\documentclass[12pt,a4paper]{article}

\usepackage[utf8]{inputenc}
\usepackage[T1]{fontenc}
\usepackage{geometry}
\usepackage{graphicx}
\usepackage{booktabs}
\usepackage{longtable}
\usepackage{amsmath,amssymb}
\usepackage{microtype}
\usepackage{hyperref}
\hypersetup{colorlinks=true,linkcolor=blue,citecolor=blue,urlcolor=blue}
\usepackage{natbib}
\usepackage{float}
\usepackage{caption}
\usepackage{xcolor}
\usepackage{enumitem}
\usepackage{csvsimple}
\usepackage{pdflscape}

\title{\textbf{Economic Corridor Detection from Multi-Source\\Earth Observation and Commercial Geospatial Data\\Across Nigeria's Kilometre-Scale Grid}\\[0.5em]
\large\normalfont Working Paper}
\author{Richard Oduntan\\
Hamoye Foundation / OpenHamoye Project\\
\texttt{richard@hamoye.org}\\[0.3em]
\small Correspondence: \texttt{research@hamoye.org}}
\date{2026}

\begin{document}
\maketitle

\begin{center}
\small\itshape
For correspondence regarding OpenHamoye data, reproducibility assets, or data-access requests,
contact \texttt{research@hamoye.org}.
\end{center}

\begin{abstract}
Identifying economic corridors at national scale is difficult in settings where firm registries, transaction flows, and local economic-output statistics are incomplete or unavailable. This paper develops a reproducible multi-proxy screening framework for detecting economic-corridor candidates across Nigeria using a kilometre-scale analytical grid. The framework fuses built-up land cover (ESA WorldCover), night-time light radiance (NASA VIIRS Black Marble), building-footprint density (Google Open Buildings), a static urban-intensity proxy derived from population, built-up share, and night-time light level, and commercial point-of-interest density from Overture Places and OpenStreetMap. A dual proxy gate requiring both built-up evidence (built-up fraction above 0.1) and night-time luminosity (radiance above 1~nW\,cm$^{-2}$\,sr$^{-1}$) identifies \emph{active corridor cells}, which are clustered into corridor objects and classified into interpretable typologies. The analytical frame contains 762,751 cells; national, state, and geopolitical-zone summaries use the 749,191 canonical national-grid cells, while 13,560 configured metropolitan area-of-interest (AOI) overlay cells are used only for an eight-metro comparison. Applied to Nigeria, the framework identifies 11,243 active corridor cells on the national grid (1.50\%), 598 corridor objects spanning 10,682 retained cells and approximately 36.3 million people, and substantial variation across metropolitan and geopolitical regions. Corridor activity concentrates around Lagos and other southern commercial geographies, and the framework also surfaces emerging and constrained corridor candidates for review. All corridor outputs are proxy-based screening signals intended for spatial planning, infrastructure prioritisation, and market-intelligence triage; they do not measure actual GDP, firm revenue, transaction volume, freight movement, infrastructure capacity, service adequacy, or official corridor designation.

\end{abstract}

\newpage
\tableofcontents
\newpage

\section{Introduction}
\label{sec:introduction}

Economic corridors --- contiguous zones of elevated commercial activity, infrastructure investment, and population density --- are fundamental units of spatial economic organisation \citep{brunner2013corridors}. In sub-Saharan Africa, where administrative statistics on firm locations and economic output are sparse, remote-sensing proxies offer an alternative pathway to corridor detection \citep{henderson2012measuring}. Night-time light radiance has been widely used as a correlate of economic activity at subnational scales \citep{chen2011using}, while built-up land cover from optical satellites provides complementary evidence of physical urbanisation \citep{pesaresi2016operating}.

Despite these advances, most corridor studies rely on a single proxy (typically night-time lights alone) and operate at coarse administrative boundaries rather than at the fine-grained cell level useful for planning and market-intelligence triage. The emergence of open building-footprint datasets \citep{sirko2021continental} and crowd-sourced business listings through Overture Places and OpenStreetMap creates an opportunity to fuse multiple proxies into a richer corridor signal, while composite-indicator practice cautions that such fused scores are interpretive screening tools rather than measured quantities \citep{oecd2008handbook}.

This paper makes four contributions:

\begin{enumerate}[leftmargin=2em]
  \item \textbf{National cell-level corridor screening.} We compute a reproducible multi-proxy corridor score over Nigeria's kilometre-scale analytical grid and identify 11,243 active corridor cells on the canonical national-grid subset through a dual proxy gate requiring both built-up fraction above 0.1 and night-time light radiance above 1~nW\,cm$^{-2}$\,sr$^{-1}$.
  \item \textbf{Corridor object construction.} We cluster adjacent active cells into 598 corridor objects spanning 10,682 cells and approximately 36.3 million people, and classify them into interpretable typologies: Mature Urban Cores, Growth Corridors, Commercial Axes, Emerging Corridors, and Constrained Corridors, plus a residual Minor Cluster category.
  \item \textbf{Multi-source geospatial fusion.} We integrate built-up land cover, night-time lights, building-footprint density, a static urban-intensity proxy, and commercial point-of-interest density, with facility point-presence context from mapped health facilities and schools.
  \item \textbf{Metro and regional comparison.} We compare corridor density across eight configured metropolitan AOIs, 36 states and the Federal Capital Territory, and six geopolitical zones.
\end{enumerate}

All corridor outputs are proxy-based screening signals. They do not measure actual GDP, firm revenue, transaction volume, freight movement, infrastructure capacity, service adequacy, or official corridor designation. The framework is intended as a transparent, reproducible first pass that surfaces corridor candidates for downstream validation against external economic records. Public-safe derived tables, methodology notes, and index definitions are released as reproducibility assets (Section~\ref{sec:availability}).

\section{Related Work}
\label{sec:related_work}

\subsection{Nighttime Lights as Economic Proxies}

The use of nighttime light (NTL) radiance as a proxy for economic activity has a two-decade lineage. \citet{henderson2012measuring} demonstrated strong correlations between DMSP-OLS luminosity and national GDP, while \citet{chen2011using} extended the approach to subnational GDP estimation. The transition from DMSP-OLS to the VIIRS Day--Night Band brought substantial improvements in spatial resolution (500\,m versus 2.7\,km) and radiometric sensitivity \citep{elvidge2017viirs}, making cell-level economic corridor detection feasible for the first time.

\subsection{Built-Up Land Cover and Urban Extent}

The ESA WorldCover product provides global land-cover classification at 10\,m resolution \citep{zanaga2021esa}, enabling precise quantification of built-up fractions within arbitrary spatial units. \citet{pesaresi2016operating} established the Global Human Settlement Layer as a complementary approach, but ESA WorldCover enables current built-up fraction measurement at high spatial resolution. Temporal built-up growth-rate computation is a future extension and is not used in the current Corridor Score, which instead uses a static Urban Intensity Index.

\subsection{Building Footprints and SMB Density}

Open building footprint datasets from Google \citep{sirko2021continental} and Microsoft have reached continental coverage, providing a census-like count of structures at individual-footprint resolution. Merchant location data from OpenStreetMap and the Overture Maps Foundation now provide a demand-side signal on commercial establishment density, though coverage remains uneven across Nigerian states.

\subsection{Economic Corridors and Composite Screening Indices}

Economic corridors are a long-standing organising concept in economic geography and regional development, linking commercial concentration, infrastructure investment, and settlement along contiguous axes \citep{brunner2013corridors}. Several studies have combined two or more remote-sensing products for urban characterisation \citep{florczyk2019description}, but cell-level multi-proxy corridor scoring at national scale --- integrating land cover, night-time lights, building counts, an urban-intensity proxy, and commercial listings --- remains underexplored. Where multiple indicators are combined into a single score, established composite-indicator guidance emphasises transparent weighting, normalisation choices, and explicit treatment of the score as an interpretive screening device rather than a measured outcome \citep{oecd2008handbook}. The present framework follows this practice: it builds on a reproducible kilometre-scale analytical grid and previously computed built-up, night-time light, population, and commercial layers, and fuses them into a transparent corridor score whose construction and index definitions are released as reproducibility assets (Section~\ref{sec:availability}).

\section{Study Area}
\label{sec:study_area}

The study covers the entirety of Nigeria, represented through a kilometre-scale analytical grid (0.01\textdegree{} spacing, $\sim$1.23\,km\textsuperscript{2} per cell). The full analytical frame contains 762,751 cells. For national, state, and geopolitical-zone summaries, we use the 749,191 canonical national-grid cells. The remaining 13,560 cells are configured metropolitan area-of-interest (AOI) overlays used only for the eight-metro comparison and are not added to state or zone denominators. Throughout the paper we therefore distinguish the ``762,751-cell analytical frame'' from the ``749,191 national-grid subset''; corridor coverage rates quoted for the nation, states, and zones use the national-grid subset as their denominator.

\medskip
\noindent The grid is organised hierarchically into:

\begin{itemize}[leftmargin=2em]
  \item \textbf{National grid:} 749,191 canonical cells across all 36 states and the FCT.
  \item \textbf{Six geopolitical zones:} South West, South East, South South, North West, North Central, and North East.
  \item \textbf{Eight metropolitan areas of interest (AOIs):} Lagos (3,318 cells), Abuja (4,636 cells), Kano (1,656 cells), Ibadan (1,271 cells), Port Harcourt (806 cells), Kaduna (676 cells), Enugu (651 cells), and Benin City (546 cells).
\end{itemize}

Nigeria's economic geography exhibits pronounced spatial heterogeneity. The south-western axis anchored by Lagos exhibits a disproportionate concentration of built-up land cover and night-time luminosity, while the northern zones are characterised by lower built-up densities and weaker night-time light signatures. This gradient makes Nigeria a useful testbed for multi-proxy corridor detection: a method must be sensitive enough to surface nascent corridor candidates in the north while remaining informative in the saturated built-up core of Lagos.

On the 749,191-cell national-grid subset, 11,243 cells (1.50\%) pass the dual proxy gate to qualify as active corridor cells; a further 8,680 cells (1.16\%) exhibit built-up--and--commercial-listing presence without meeting the night-time light threshold. Across the wider analytical frame, built-up land cover is present in roughly 3.9\% of cells. These proxy shares describe the screening signal and are not measures of economic output.

\section{Data}
\label{sec:data}

Table~\ref{tab:dataset_inventory} summarises the datasets underpinning the corridor analysis. All raster and vector sources are ingested through reproducible extract--transform--load pipelines and aggregated to the kilometre-scale analytical grid (0.01\textdegree{} spacing, $\sim$1.11\,km $\times$ 1.11\,km).

\begin{table}[H]
\centering
\caption{Corridor Dataset Inventory}
\label{tab:dataset_inventory}
\small
\begin{tabular}{llllp{4cm}}
\toprule
\textbf{Dataset} & \textbf{Source} & \textbf{Res.} & \textbf{Epoch} & \textbf{Role} \\
\midrule
ESA WorldCover & ESA / Copernicus & 10\,m & 2021 & Built-up fraction \\
VIIRS Black Marble & NASA & 500\,m & 2023--2024 & NTL radiance \\
Google Open Buildings & Google Research & Footprints & 2024 & Building count \\
WorldPop Density & WorldPop & 100\,m & 2020 & Population context \\
Overture / OSM & Overture / OSM & Point & 2024 & SMB count \\
Urban Intensity Index & Derived & Cell & Derived & Static urban-intensity proxy \\
Infrastructure-stress proxy & Derived & Cell & Derived & Accessibility/service proxy \\
Hazard-burden proxy (HBS) & Derived & Cell & Derived & Composite hazard-burden proxy \\
Admin Boundaries & geoBoundaries & Polygons & Static & State/LGA/metro \\
\bottomrule
\end{tabular}
\end{table}

\subsection{Data Quality Notes}

Night-time light data is now available for all eight metro AOIs following national VIIRS propagation. The metro corridor density comparison (Table~\ref{tab:metro_density}) reflects the current nationally complete NTL coverage.

SMB merchant counts are available for national-grid cells but not yet propagated to metro-AOI cells, resulting in zero SMB counts across all eight metro AOIs. The SMB component thus contributes to corridor scores only for national-grid cells. Because SMB is zero in metro-AOI cells, score-based comparisons between national-grid and metro-AOI cells should be interpreted cautiously. Metro corridor-density comparisons are based on the dual proxy gate and are not directly affected by the SMB component.

Google Open Buildings provides building-density context for 479,088 cells nationally (62.8\% building-positive cell share, not source coverage). GRID3 health facilities (46,146) and schools (204,513) provide facility point-presence context for all corridors. These are point-containing cell shares, not service-access coverage.

\medskip
\noindent The source datasets are: ESA WorldCover \citep{zanaga2021esa}, NASA VIIRS Black Marble \citep{roman2018viirs}, Google Open Buildings \citep{sirko2021continental}, WorldPop \citep{tatem2017worldpop}, Overture Maps \citep{overture2024} and OpenStreetMap \citep{openstreetmap}, GRID3 facilities \citep{grid3}, and geoBoundaries administrative boundaries \citep{runfola2020geoboundaries}. Each remains governed by its original licence; only derived, public-safe outputs are released (Section~\ref{sec:availability}).

\section{Methodology}
\label{sec:methodology}

\subsection{Corridor Proxy Gate}

A cell qualifies as an \emph{active corridor cell} if and only if it satisfies both conditions of the dual proxy gate:

\begin{equation}
\text{gate}(c) = \mathbb{1}\!\left[\text{builtup}(c) > 0.1\right] \wedge \mathbb{1}\!\left[\text{nl}(c) > 1\right]
\label{eq:proxy_gate}
\end{equation}

\noindent where $\text{builtup}(c)$ is the ESA WorldCover built-up fraction and $\text{nl}(c)$ is the mean VIIRS radiance in nW\,cm$^{-2}$\,sr$^{-1}$. The built-up threshold captures physical urbanisation, while the night-time light threshold captures nocturnal economic or electrified activity; requiring both reduces false positives from bare built-up land with no luminosity and from transient light with no settlement. Both thresholds are heuristic and interpretable rather than statistically optimised; sensitivity analysis over alternative thresholds is left to future work. Applying this gate to the canonical national-grid subset (749,191 unique cells, excluding 13,560 configured metro AOI overlay cells) yields 11,243 active corridor cells (1.50\%). The broader 762,751-cell analytical frame includes the metro AOI overlays; these are analysed separately for the eight-metro comparison and are not added to state or zone denominators.

\subsection{Corridor Score}

For cells passing the proxy gate, the Corridor Score is a weighted linear combination of five normalised components:

\begin{equation}
\begin{aligned}
\text{score}(c) = {}& 0.30 \times \text{builtup}(c) + 0.25 \times \hat{\text{nl}}(c) + 0.20 \times \text{UII}(c) \\
& + 0.15 \times \hat{\text{bldg}}(c) + 0.10 \times \hat{\text{smb}}(c)
\end{aligned}
\label{eq:corridor_score}
\end{equation}

\noindent where $\hat{\text{nl}}(c) = \min(\text{nl}/50, 1)$, $\hat{\text{bldg}}(c) = \min(\text{bldg}/3000, 1)$, and $\hat{\text{smb}}(c) = \min(\text{smb}/100, 1)$. The Urban Intensity Index (UII) is a static composite of normalised population density (weight 0.4), built-up share (0.3), and night-time light level (0.3); it is a cross-sectional intensity proxy and \textbf{not} a temporal growth rate. The weights above are set by expert judgement rather than empirical calibration against a ground-truth economic-output measure (Section~\ref{sec:limitations}), and the corridor score should be read as a transparent proxy-based screening score.

\begin{table}[H]
\centering
\caption{Corridor Index Component Definitions}
\label{tab:index_defs}
\small
\begin{tabular}{lllp{5cm}}
\toprule
\textbf{Component} & \textbf{Weight} & \textbf{Source} & \textbf{Description} \\
\midrule
builtup & 0.30 & ESA WorldCover & Built-up fraction (0--1) \\
nl\_norm & 0.25 & NASA VIIRS & $\min(\text{nl}/50, 1)$: normalised radiance \\
UII & 0.20 & Derived & Static urban-intensity proxy \\
bldg\_norm & 0.15 & Google Open Buildings & $\min(\text{bldg}/3000, 1)$ \\
smb\_norm & 0.10 & Overture / OSM & $\min(\text{smb}/100, 1)$ \\
\bottomrule
\end{tabular}
\end{table}

\subsection{Corridor Object Construction}

Adjacent active corridor cells are clustered using queen adjacency (8-connected) on the kilometre-scale grid, retaining clusters of $\geq 3$ cells. This yields 598 corridor objects.

\subsection{Corridor Typology}

Corridor objects are classified using heuristic thresholds applied in the following precedence order (first match wins):
\begin{enumerate}[leftmargin=2em]
  \item \textbf{Mature Urban Core}: $\geq 50$ cells AND mean NTL $> 10$
  \item \textbf{Growth Corridor}: $\geq 20$ cells AND mean UII $> 0.1$
  \item \textbf{Commercial Axis}: $\geq 10$ cells AND mean builtup $> 0.3$
  \item \textbf{Constrained Corridor}: mean HBS $> 0.1$
  \item \textbf{Emerging Corridor}: mean UII $> 0.05$
  \item \textbf{Minor Cluster}: residual (does not satisfy any substantive typology)
\end{enumerate}

\noindent Five substantive typologies account for 587 objects; 11 additional objects are retained as Minor Clusters. The typology uses heuristic, first-match thresholds and is intended to organise corridor objects for interpretation. It is not an official economic classification and should not be read as a verified planning designation: ``Constrained Corridor'' does not imply a proven infrastructure deficit, ``Emerging Corridor'' does not imply confirmed future growth, ``Commercial Axis'' does not imply verified business revenue, and ``Mature Urban Core'' does not imply an official planning status. The denominator caps (NTL/50, bldg/3000, SMB/100) are heuristic normalisation thresholds; sensitivity analysis over alternative caps is future work.

\subsection{Constrained and Emerging Corridor Detection}

Constrained corridor candidates are active corridor cells or objects where HBS exceeds 0.1, indicating elevated hazard-burden proxy values. They are screening candidates, not confirmed infrastructure deficits.

Emerging corridor candidates are cells that fail the dual corridor gate but exhibit high static UII and market-potential signals. In the current implementation, candidates are selected from cells where $\text{gate}(c) = 0$ and ranked by a composite of UII and market potential; the top-20 table reports the highest-scoring candidates rather than a threshold-defined class.

All corridor scores are proxy-based screening tools and do not measure actual economic output, firm-level revenue, or infrastructure capacity.

\section{Results}
\label{sec:results}

\begin{figure}[H]
\centering
\includegraphics[width=\textwidth]{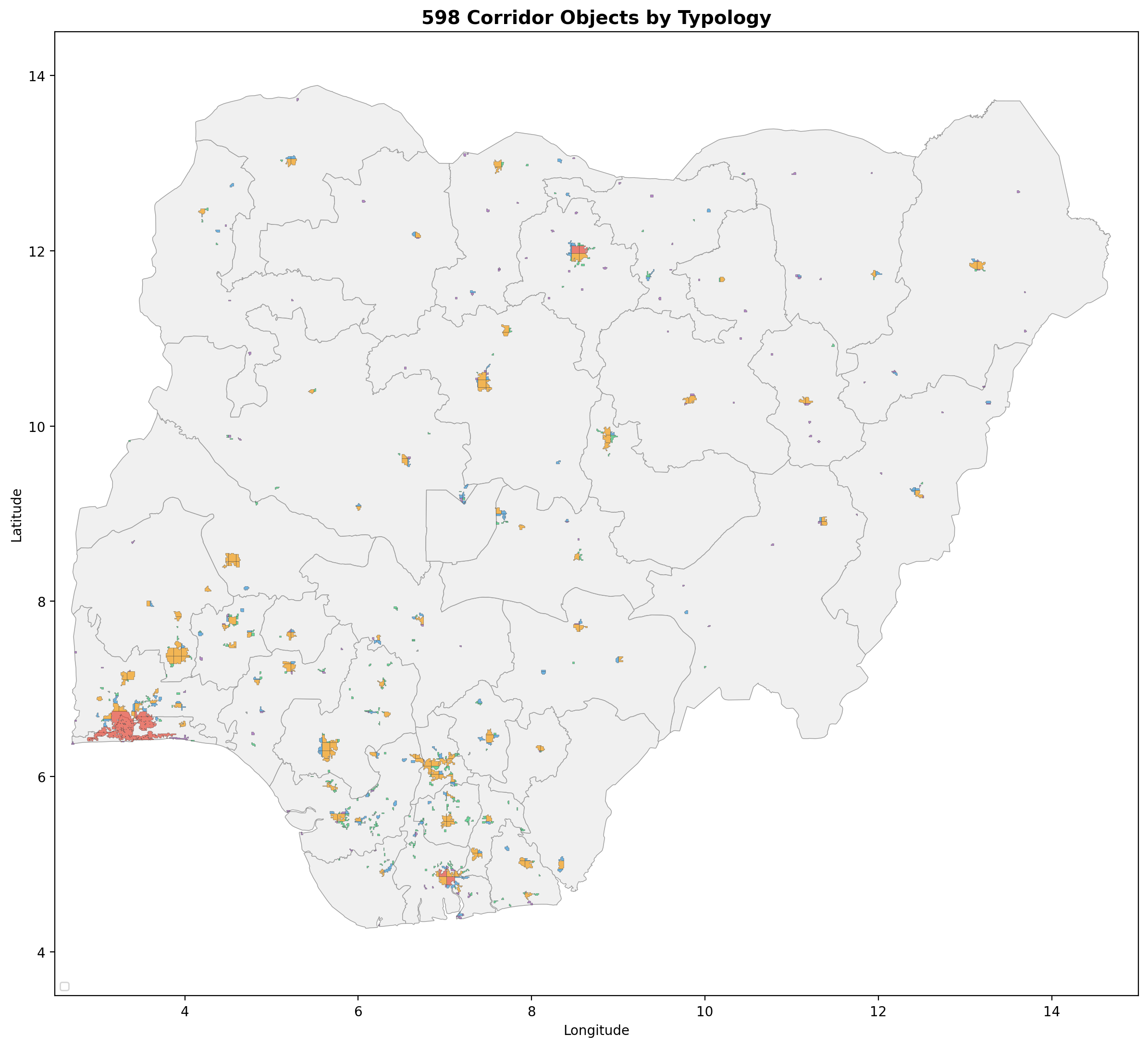}
\caption{598 corridor objects by typology across Nigeria. Mature Urban Cores (red) concentrate in Lagos and Kano; Growth Corridors (orange) are distributed across southern and central Nigeria.}
\label{fig:corridor_map}
\end{figure}

\begin{figure}[H]
\centering
\includegraphics[width=\textwidth]{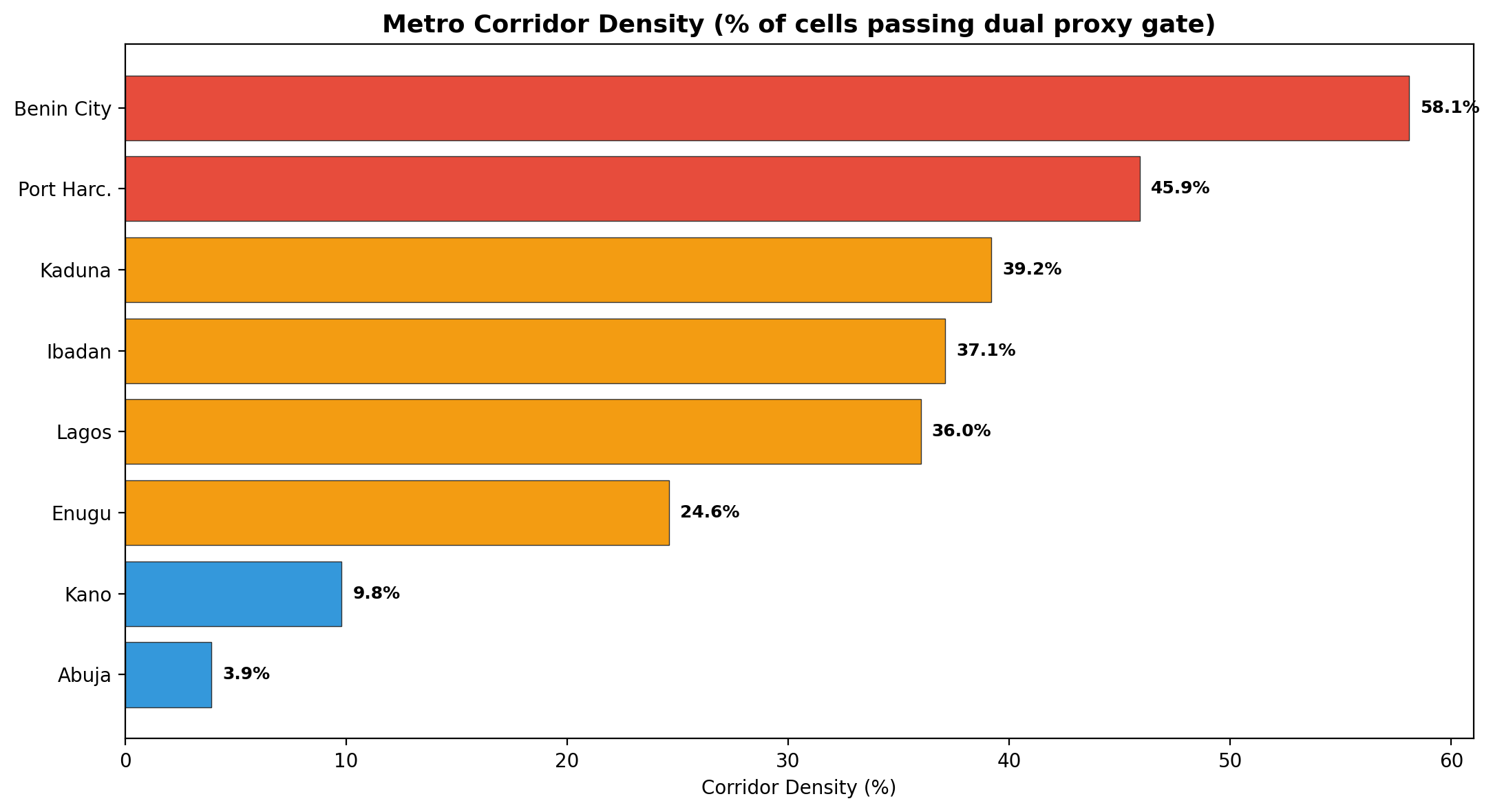}
\caption{Metropolitan corridor density. Benin City leads at 58.1\%; Abuja trails at 3.9\% due to its large dispersed metro footprint.}
\label{fig:metro_density}
\end{figure}

\begin{figure}[H]
\centering
\includegraphics[width=\textwidth]{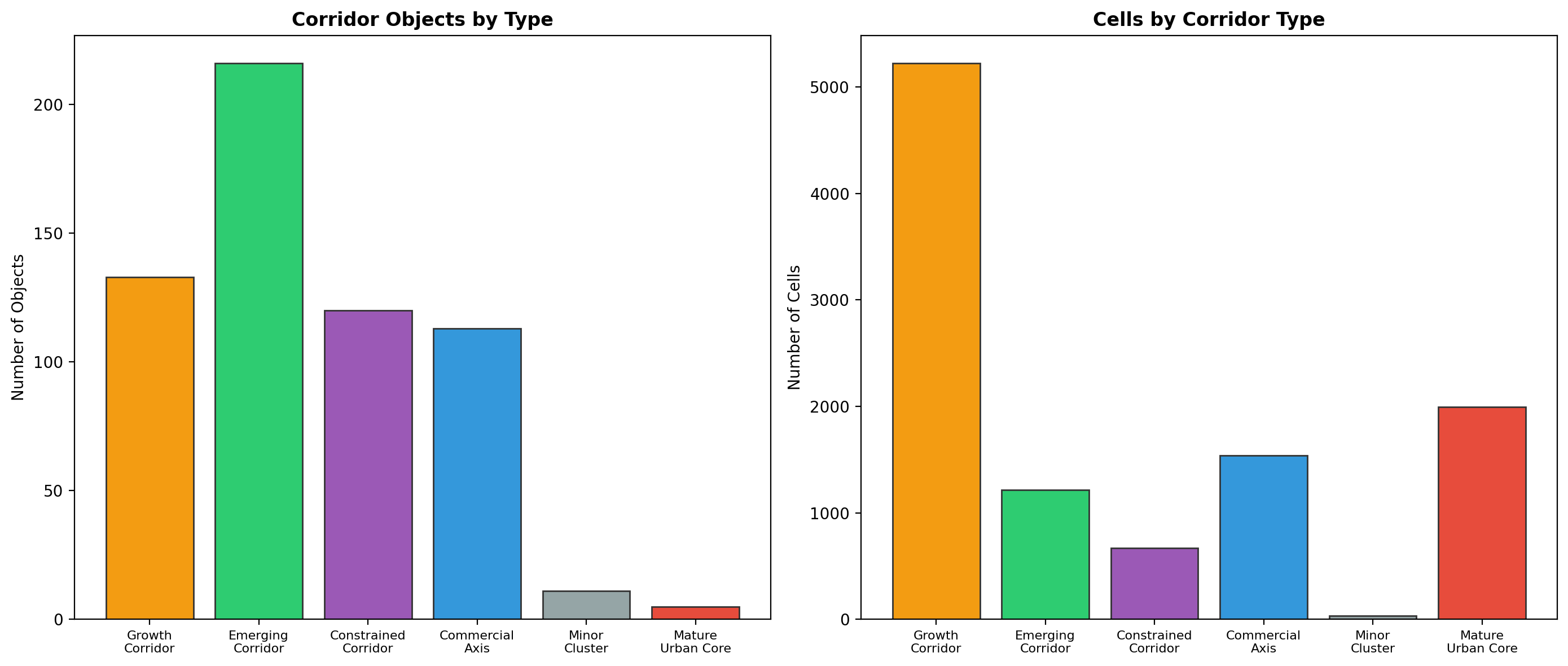}
\caption{Corridor object and cell distribution by typology. Emerging Corridors are most numerous (217 objects) but contain fewer cells; Growth Corridors account for the most cells (5,224).}
\label{fig:typology}
\end{figure}

Figure~\ref{fig:corridor_map} maps all 598 corridor objects, Figure~\ref{fig:metro_density} compares metro corridor densities, and Figure~\ref{fig:typology} shows the typology distribution.

\subsection{National Corridor Coverage}

Table~\ref{tab:coverage_audit} summarises corridor cell counts at the national level.

\begin{table}[H]
\centering
\caption{National Corridor Coverage Audit}
\label{tab:coverage_audit}
\begin{tabular}{lrr}
\toprule
\textbf{Metric} & \textbf{Count} & \textbf{Share} \\
\midrule
Total cells (national grid) & 749,191 & 100.00\% \\
Active corridor cells (gate pass) & 11,243 & 1.50\% \\
Retained in clusters ($\geq 3$) & 10,682 & 1.43\% \\
Excluded isolated/small-cluster & 561 & 0.07\% \\
Corridor objects & 598 & --- \\
Non-corridor cells & 737,948 & 98.50\% \\
\bottomrule
\end{tabular}
\end{table}

Of the 11,243 active corridor cells on the national grid, 10,682 are retained in clusters of at least three cells; the remaining 561 are isolated or belong to clusters below the minimum threshold. The national-grid subset contains 749,191 cells; the 13,560 configured metro AOI overlay cells are analysed separately for the metro comparison and are not added to the national-grid denominator. Facility point-presence context is drawn from 46,146 mapped GRID3 health facilities and 204,513 schools, and building-footprint context from Google Open Buildings (present in 479,088 cells, 62.8\% of the frame). These are point-containing and footprint-present cell shares, not measures of service access, facility capacity, or infrastructure condition.

\subsection{Top Corridor Cells}

Table~\ref{tab:top30} presents the 30 highest-scoring corridor cells nationally. The top cell is cell~995 in Lagos with a corridor score of 0.634, driven by high built-up fraction (0.943), strong night-time light radiance (91.4~nW\,cm$^{-2}$\,sr$^{-1}$), and an above-average UII of 0.506. Lagos cells account for 20 of the top 30 positions. National-grid cells from the broader Lagos conurbation and other southern cities fill the remaining positions.

\begin{table}[H]
\centering
\caption{Top 10 Corridor Cells by Score (of 30 total; see supplementary CSV for full ranking)}
\label{tab:top30}
\small
\begin{tabular}{rlrrrrrr}
\toprule
\textbf{Rank} & \textbf{Cell} & \textbf{AOI} & \textbf{Score} & \textbf{Builtup} & \textbf{NL} & \textbf{UII} & \textbf{Bldg} \\
\midrule
1 & 995 & Lagos & 0.634 & 0.943 & 91.43 & 0.506 & 1,239 \\
2 & 977 & Lagos & 0.617 & 0.966 & 53.55 & 0.386 & 2,447 \\
3 & 906 & Lagos & 0.611 & 0.983 & 35.31 & 0.695 & 3,315 \\
4 & 1161 & Lagos & 0.607 & 0.983 & 89.32 & 0.311 & 2,408 \\
5 & 981 & Lagos & 0.606 & 0.948 & 49.12 & 0.381 & 2,235 \\
6 & 1021 & Lagos & 0.605 & 0.926 & 109.81 & 0.387 & 3,153 \\
7 & 92495 & National & 0.603 & 0.976 & 53.90 & 0.301 & 2,089 \\
8 & 92494 & National & 0.598 & 0.961 & 53.98 & 0.297 & 2,122 \\
9 & 1138 & Lagos & 0.596 & 0.953 & 98.60 & 0.303 & 1,775 \\
10 & 88369 & National & 0.596 & 0.947 & 60.33 & 0.309 & 1,895 \\
\bottomrule
\end{tabular}
\end{table}

\subsection{Metro-Level Corridor Density}

With night-time light data now propagated to all eight metro AOIs, the full metro corridor comparison is available (Table~\ref{tab:metro_density}). Benin City leads with the highest corridor density (58.1\%), followed by Port Harcourt (45.9\%) and Kaduna (39.2\%). Lagos, despite having the most corridor cells in absolute terms (1,193), ranks fifth in density (36.0\%) due to its larger total metro footprint.

\begin{table}[H]
\centering
\caption{Metropolitan Corridor Density (All 8 Metros)}
\label{tab:metro_density}
\begin{tabular}{lrrrrrrr}
\toprule
\textbf{Metro} & \textbf{Total} & \textbf{Corridor} & \textbf{Pct} & \textbf{Builtup} & \textbf{NL} & \textbf{Health} & \textbf{Schools} \\
\midrule
Benin City & 546 & 317 & 58.1\% & 0.440 & 2.59 & 287 & 632 \\
Port Harcourt & 806 & 370 & 45.9\% & 0.297 & 9.87 & 238 & 1,357 \\
Kaduna & 676 & 265 & 39.2\% & 0.263 & 2.83 & 109 & 1,186 \\
Ibadan & 1,271 & 472 & 37.1\% & 0.306 & 1.98 & 766 & 2,237 \\
Lagos & 3,318 & 1,193 & 36.0\% & 0.237 & 7.01 & 2,379 & 15,745 \\
Enugu & 651 & 160 & 24.6\% & 0.171 & 1.76 & 418 & 1,585 \\
Kano & 1,656 & 162 & 9.8\% & 0.175 & 2.91 & 482 & 3,670 \\
Abuja & 4,636 & 181 & 3.9\% & 0.098 & 1.06 & 637 & 2,048 \\
\bottomrule
\end{tabular}
\end{table}

\noindent\textit{Health and Schools are GRID3 mapped facility point counts within the configured metro AOI; they do not measure service access, facility capacity, or functionality. Metro AOIs are configured analytical footprints, not official metropolitan boundaries.}

Benin City's high corridor density (58.1\%) is consistent with a compact, highly built-up urban form within its configured AOI (mean built-up 0.440), while Port Harcourt's elevated density (45.9\%) coincides with strong night-time light signatures (mean NL 9.87) associated with oil-and-gas infrastructure. Abuja's low corridor density (3.9\%) reflects its large configured metro footprint (4,636 cells) relative to its concentrated built-up core. Because these densities are computed over configured analytical footprints of differing extent, cross-metro comparisons should be read as relative screening signals rather than official metropolitan statistics.

\subsection{Corridor Object Typology}

Adjacent active corridor cells are clustered into corridor objects using queen adjacency (8-connected) on the 1\,km grid, retaining clusters of 3+ cells. This yields 598 corridor objects spanning 10,682 cells and an estimated 36.3 million people. Table~\ref{tab:corridor_types} summarises the typology distribution.

\begin{table}[H]
\centering
\caption{Corridor Object Typology (598 Objects, 10,682 Cells, 36.3M Population)}
\label{tab:corridor_types}
\begin{tabular}{lrrrr}
\toprule
\textbf{Type} & \textbf{Objects} & \textbf{Cells} & \textbf{Pop (M)} & \textbf{Avg Score} \\
\midrule
Growth Corridor & 133 & 5,224 & 14.5 & 0.29 \\
Mature Urban Core & 5 & 1,996 & 17.1 & 0.42 \\
Commercial Axis & 113 & 1,540 & 2.6 & 0.27 \\
Emerging Corridor & 217 & 1,224 & 1.6 & 0.22 \\
Constrained Corridor & 119 & 663 & 0.5 & 0.24 \\
Minor Cluster & 11 & 35 & 0.02 & 0.18 \\
\bottomrule
\end{tabular}
\end{table}

The largest corridor object is the Lagos Mature Urban Core (1,719 cells, 14.5M population, score 0.504), followed by Kano's two urban cores (80 and 69 cells). Growth Corridors are the most numerous type (133 objects), concentrated in Oyo, Kaduna, Borno, Edo, and Ogun states. Facility point-presence context from GRID3 shows that the Lagos core contains 2,942 health facilities and 20,000 schools, while many Growth Corridors have lower mapped facility point counts relative to their population.

\subsection{State-Level Corridor Density}

Lagos state leads with a corridor density of 36.63\% (2,286 corridor cells out of 6,240 total), more than double the second-ranked state, Anambra (14.63\%). Table~\ref{tab:state_density} lists the top 15 states.

\begin{table}[H]
\centering
\caption{Top 15 States by Corridor Density}
\label{tab:state_density}
\small
\begin{tabular}{rlllrr}
\toprule
\textbf{Rank} & \textbf{State} & \textbf{Zone} & \textbf{Total} & \textbf{Corridor} & \textbf{Pct} \\
\midrule
1 & Lagos & South West & 6,240 & 2,286 & 36.63\% \\
2 & Anambra & South East & 3,937 & 576 & 14.63\% \\
3 & Imo & South East & 4,159 & 380 & 9.14\% \\
4 & Ogun & South West & 13,963 & 1,200 & 8.59\% \\
5 & Rivers & South South & 8,479 & 676 & 7.97\% \\
6 & Abia & South East & 3,910 & 243 & 6.21\% \\
7 & Delta & South South & 14,000 & 762 & 5.44\% \\
8 & Osun & South West & 7,006 & 316 & 4.51\% \\
9 & Akwa Ibom & South South & 5,677 & 254 & 4.47\% \\
10 & Edo & South South & 16,004 & 598 & 3.74\% \\
11 & Enugu & South East & 6,233 & 203 & 3.26\% \\
12 & Kano & North West & 16,775 & 477 & 2.84\% \\
13 & Oyo & South West & 22,779 & 620 & 2.72\% \\
14 & Ondo & South West & 12,344 & 264 & 2.14\% \\
15 & Bayelsa & South South & 7,775 & 142 & 1.83\% \\
\bottomrule
\end{tabular}
\end{table}

\subsection{Geopolitical Zone Summary}

The south--north gradient in corridor density is stark (Table~\ref{tab:zone_summary}). The South East leads nationally at 6.19\%, followed by the South West (5.59\%) and South South (3.65\%). Northern zones trail significantly: North West at 0.85\%, North Central at 0.74\%, and North East at 0.36\%.

\begin{table}[H]
\centering
\caption{Geopolitical Zone Corridor Summary}
\label{tab:zone_summary}
\begin{tabular}{lrrrr}
\toprule
\textbf{Zone} & \textbf{States} & \textbf{Total} & \textbf{Corridor} & \textbf{Pct} \\
\midrule
South East & 5 & 23,546 & 1,457 & 6.19\% \\
South West & 6 & 63,464 & 3,549 & 5.59\% \\
South South & 6 & 69,248 & 2,527 & 3.65\% \\
North West & 7 & 179,381 & 1,531 & 0.85\% \\
North Central & 7 & 177,223 & 1,308 & 0.74\% \\
North East & 6 & 234,854 & 844 & 0.36\% \\
\bottomrule
\end{tabular}
\end{table}

\noindent\textit{Note: Zone summaries are computed from unique national-grid cells only. Metro AOI cells are not added to state/zone totals. Zone totals sum to 747,716 cells; the remaining 1,475 national-grid cells lack zone assignment due to boundary-edge geometry or null LGA attribution and are excluded from the six-zone denominator. Metro AOIs are configured analytical footprints, not official metropolitan boundaries.}

\section{Discussion}
\label{sec:discussion}

\subsection{Lagos Concentration and the Primacy Pattern}

The finding that Lagos accounts for 1,193 of the metro-level corridor cells and 2,286 of state-level corridor cells (36.63\% density) is consistent with Nigeria's well-documented primate-city pattern. Cell~995, the top-scoring corridor nationally (score $= 0.634$), combines near-complete built-up coverage (0.943), high night-time light radiance (91.43~nW\,cm$^{-2}$\,sr$^{-1}$), and above-average UII (0.506), placing it in the high-density built-up core of Lagos Island and its environs. The presence of 20 of the top 30 corridor cells in Lagos indicates a pronounced concentration of the corridor screening signal in Lagos relative to other Nigerian metros. This is a spatial concentration of the built-up and luminosity proxies, not a measure of Lagos's share of national output.

\subsection{The South--North Gradient}

Corridor density on the national-grid subset falls by roughly seventeen-fold between the South East (6.19\%) and the North East (0.36\%), consistent with broad differences in urbanisation intensity, night-time luminosity, and built-up concentration rather than a direct measure of investment or infrastructure gaps. The South East leads (6.19\%), driven by Anambra (14.63\%) and Imo (9.14\%); the South West follows (5.59\%), anchored by Lagos. Kano is the sole northern state in the top-15 corridor-density ranking, consistent with its role as the principal commercial hub of northern Nigeria.

\subsection{Corridor Score Decomposition}

The corridor score's five components contribute unevenly across geographies. In Lagos, built-up fraction is near-saturated (most top cells exceed 0.92), so night-time light intensity and UII become the primary differentiators. In national-grid cells outside Lagos, SMB merchant density provides additional discrimination: cell~88370 near Lagos's national-grid boundary has 697 SMB merchants, the highest in the top 30, contributing meaningfully to its score of 0.582.

The infrastructure index (infra) and HBS / hazard-burden proxy do not enter the corridor score directly but serve as diagnostic layers. Top corridor cells in Lagos show infrastructure scores ranging from 0.30 to 0.75, reflecting heterogeneous accessibility even within the economic core.

\subsection{Policy Implications}

The corridor map enables spatially targeted policy interventions:
\begin{itemize}[leftmargin=2em]
  \item \textbf{Infrastructure investment:} Constrained corridors (Section~\ref{sec:constrained}) identify cells where economic activity is already present but infrastructure-stress proxy values may warrant review.
  \item \textbf{Market development:} Emerging corridors (Section~\ref{sec:emerging}) highlight cells with high static urban-intensity and market-potential signals where infrastructure and service-planning screening may be warranted.
  \item \textbf{Financial inclusion:} The south--north corridor density gradient can guide bank branch placement, mobile money agent networks, and microfinance allocation strategies.
\end{itemize}

\section{Constrained Corridors}
\label{sec:constrained}

Constrained corridor candidates are active corridor cells where corridor signals coincide with elevated HBS and/or infrastructure-stress proxy values. They should be interpreted as screening candidates for hazard- or infrastructure-stress review, not as direct measures of engineered infrastructure deficits or economic underperformance. Table~\ref{tab:constrained} lists the 20 most constrained corridor cells.

\begin{table}[H]
\centering
\caption{Top 20 Constrained Corridor Candidate Cells Ranked by HBS}
\label{tab:constrained}
\small
\begin{tabular}{rlrrrrr}
\toprule
\textbf{Rank} & \textbf{Cell} & \textbf{AOI} & \textbf{HBS} & \textbf{Infra} & \textbf{Density} & \textbf{Mkt Pot} \\
\midrule
1 & 1460 & Lagos & 0.360 & 0.428 & -- & -- \\
2 & 89735 & National & 0.359 & 0.385 & 7,277 & 0.292 \\
3 & 1090 & Lagos & 0.357 & 0.430 & 3,178 & -- \\
4 & 89046 & National & 0.356 & 0.380 & 7,277 & 0.277 \\
5 & 89390 & National & 0.355 & 0.388 & 7,277 & 0.302 \\
6 & 905 & Lagos & 0.354 & 0.446 & 7,277 & -- \\
7 & 384079 & National & 0.352 & 0.366 & 2,022 & 0.254 \\
8 & 1067 & Lagos & 0.351 & 0.423 & 7,277 & -- \\
9 & 90080 & National & 0.350 & 0.388 & 3,356 & 0.307 \\
10 & 1018 & Lagos & 0.350 & 0.440 & 7,277 & -- \\
11 & 1091 & Lagos & 0.348 & 0.416 & 3,178 & -- \\
12 & 221977 & National & 0.348 & 0.364 & 181 & 0.247 \\
13 & 990 & Lagos & 0.345 & 0.458 & 7,277 & -- \\
14 & 1270 & Lagos & 0.343 & 0.414 & -- & -- \\
15 & 384080 & National & 0.341 & 0.401 & 2,022 & 0.356 \\
16 & 1043 & Lagos & 0.341 & 0.473 & 11,331 & -- \\
17 & 89392 & National & 0.341 & 0.401 & 5,230 & 0.339 \\
18 & 92151 & National & 0.341 & 0.362 & 1,472 & 0.242 \\
19 & 89736 & National & 0.340 & 0.409 & 13,597 & 0.343 \\
20 & 90084 & National & 0.338 & 0.363 & 1,472 & 0.244 \\
\bottomrule
\end{tabular}
\end{table}

\subsection{Patterns in Constrained Corridors}

Several patterns emerge:

\begin{enumerate}[leftmargin=2em]
  \item \textbf{Lagos periphery:} Ten of the 20 constrained cells are in Lagos, predominantly on the metro's flood-prone edges (cells 1460, 1090, 905, 1067, 1018, 1091, 990, 1270, 1043). These cells exhibit infrastructure scores in the 0.41--0.47 range, below the top-corridor median.
  \item \textbf{National-grid constraints:} The remaining 10 cells are scattered across the national grid, with population densities ranging from 181 to 13,597 persons per km$^{2}$. Cell~221977 stands out with extremely low density (181), suggesting a remote but economically active node --- possibly a mining or agricultural processing site.
  \item \textbf{Market potential gap:} National-grid constrained cells show market potential scores between 0.24 and 0.36, well below the top-corridor average of approximately 0.51. This suggests that constrained-corridor candidates may combine corridor activity signals with hazard-burden or infrastructure-stress concerns requiring local validation.
\end{enumerate}

\section{Emerging Corridors}
\label{sec:emerging}

Emerging corridor candidates are cells that do not currently pass the dual proxy gate but exhibit high static UII and market-potential signals. Zero or low night-time light signatures indicate that these cells remain below the VIIRS luminosity threshold; they should not be interpreted as lacking all electrification or commercial activity. Table~\ref{tab:emerging} lists the 20 top-ranked emerging corridor candidates.

\begin{table}[H]
\centering
\caption{Top 20 Emerging Corridor Candidate Cells Ranked by UII + Market-Potential Score}
\label{tab:emerging}
\small
\begin{tabular}{rlrrrrr}
\toprule
\textbf{Rank} & \textbf{Cell} & \textbf{AOI} & \textbf{UII} & \textbf{Builtup} & \textbf{Density} & \textbf{Mkt Pot} \\
\midrule
1 & 829644 & Benin City & 0.400 & 0.150 & 7,049 & 0.725 \\
2 & 829530 & Benin City & 0.400 & 0.187 & 7,049 & 0.725 \\
3 & 829532 & Benin City & 0.400 & 0.052 & 7,049 & 0.725 \\
4 & 829509 & Benin City & 0.400 & 0.052 & 7,049 & 0.725 \\
5 & 829727 & Benin City & 0.400 & 0.058 & 7,049 & 0.725 \\
6 & 828378 & Kaduna & 0.400 & 0.280 & 5,664 & 0.725 \\
7 & 829529 & Benin City & 0.400 & 0.281 & 7,049 & 0.725 \\
8 & 829554 & Benin City & 0.400 & 0.253 & 7,049 & 0.725 \\
9 & 829531 & Benin City & 0.400 & 0.194 & 7,049 & 0.725 \\
10 & 829534 & Benin City & 0.400 & 0.068 & 7,049 & 0.725 \\
11 & 829728 & Benin City & 0.400 & 0.066 & 7,049 & 0.725 \\
12 & 829504 & Benin City & 0.385 & 0.187 & 6,825 & 0.708 \\
13 & 829545 & Benin City & 0.385 & 0.063 & 6,825 & 0.708 \\
14 & 829567 & Benin City & 0.385 & 0.167 & 6,825 & 0.708 \\
15 & 829483 & Benin City & 0.385 & 0.193 & 6,825 & 0.708 \\
16 & 829525 & Benin City & 0.385 & 0.103 & 6,825 & 0.708 \\
17 & 829546 & Benin City & 0.385 & 0.103 & 6,825 & 0.708 \\
18 & 829591 & Benin City & 0.385 & 0.280 & 6,825 & 0.708 \\
19 & 828497 & Kaduna & 0.267 & 0.190 & 4,399 & 0.575 \\
20 & 828440 & Kaduna & 0.267 & 0.260 & 4,399 & 0.575 \\
\bottomrule
\end{tabular}
\end{table}

\subsection{Geographic Concentration}

The emerging corridor candidate list is concentrated in two cities:

\begin{itemize}[leftmargin=2em]
  \item \textbf{Benin City (17 of 20 cells):} Cells in the Benin City metro periphery show uniformly high UIIs (0.385--0.400), moderate population densities (6,825--7,049 persons/km$^{2}$), and the highest market potential scores in the dataset (0.708--0.725). Built-up fractions range widely from 0.05 to 0.28, indicating a mix of early-stage and mid-stage urbanisation. All 17 cells have zero night-time light radiance, placing them below the corridor proxy gate despite strong static urban-intensity signal.
  \item \textbf{Kaduna (3 of 20 cells):} Two cells (828497, 828440) show moderate UII (0.267) and market potential (0.575), while cell 828378 matches the Benin City top tier with 0.400 UII and 0.725 market potential. Kaduna's emerging-corridor candidates show static urban-intensity and market-potential signals on the configured metro fringe.
\end{itemize}

\subsection{Implications for Corridor Formation}

These 20 cells are candidate emerging-corridor geographies under the current static proxy model. Their zero or low night-time light signatures indicate that these cells remain below the VIIRS luminosity threshold used by the corridor gate; this should not be interpreted as absence of all electrification or commercial activity. Targeted interventions --- grid extension, road improvements, microfinance access --- could be evaluated as part of local infrastructure, service-delivery, and market-development screening.

\section{Limitations and Future Work}
\label{sec:limitations}

The framework is a proxy-based screening tool, and its limitations should be read alongside every result reported above.

\subsection{Interpretation and Validation}

\begin{enumerate}[leftmargin=2em]
  \item \textbf{No ground-truth economic-output validation.} The corridor score is not calibrated against GDP, tax records, firm revenue, freight or transport flows, or official business-register data. It is a composite of physical and activity proxies, and a high score indicates a strong proxy signal, not measured economic output.
  \item \textbf{Proxy thresholds are heuristic.} The built-up ($>0.1$) and night-time light ($>1$~nW\,cm$^{-2}$\,sr$^{-1}$) gate thresholds are interpretable but not statistically optimised; sensitivity analysis across threshold ranges would strengthen robustness claims.
  \item \textbf{Weights are expert-set.} The five corridor-score weights (0.30, 0.25, 0.20, 0.15, 0.10) and the linear (additively separable) combination reflect expert judgement rather than empirical calibration; non-linear interactions among components are not captured.
\end{enumerate}

\subsection{Data}

\begin{enumerate}[leftmargin=2em, resume]
  \item \textbf{Commercial-listing coverage is uneven.} Overture Places and OpenStreetMap merchant listings are biased toward Lagos and southern cities and are not a census of businesses; the commercial point-of-interest component (weight 0.10) consequently has minimal impact outside Lagos and is zero in the configured metro-AOI cells.
  \item \textbf{Facility counts are point-presence, not access or capacity.} GRID3 health-facility (46,146) and school (204,513) counts are point-containing cell shares; they do not measure staffing, functionality, service quality, capacity, affordability, or actual access.
  \item \textbf{Metro AOIs are analytical footprints.} The eight metropolitan areas of interest are configured analytical footprints of differing extent, not official metropolitan boundaries; cross-metro densities are relative screening signals.
  \item \textbf{Static cross-sectional snapshot.} The corridor score reflects a single epoch (ESA WorldCover 2021, VIIRS 2023--2024). It does not estimate corridor formation, expansion, or contraction dynamics.
\end{enumerate}

\subsection{Method}

\begin{enumerate}[leftmargin=2em, resume]
  \item \textbf{Corridor objects are grid-derived.} Object boundaries and counts depend on cell resolution, the queen-adjacency (8-connected) rule, and the minimum cluster size ($\geq 3$ cells); different choices would yield different objects and typologies.
  \item \textbf{Night-time-light gate bias.} The gate may exclude low-light emerging clusters that remain below VIIRS visibility thresholds; low luminosity may reflect sensor visibility limits rather than the absence of activity.
\end{enumerate}

\subsection{Data Coverage Context}

The current analysis draws on nationally propagated inputs: night-time lights across all eight configured metro AOIs (enabling the full metro comparison); Google Open Buildings present in 479,088 cells (62.8\%), with metro coverage of roughly 82--99\%; GRID3 health facilities (46,146 across 35,569 cells) and schools (204,513 across 60,951 cells) spanning all 37 states and the FCT; and composite hazard-burden proxy values for 761,359 cells (99.82\%). These describe input availability, not validated economic or service outcomes.

\subsection{Future Work}

\begin{itemize}[leftmargin=2em]
  \item Calibrate corridor-score weights and gate thresholds against external economic records (firm revenue, tax, or business-register data) where available.
  \item Extend temporal coverage to construct corridor formation and decay trajectories, and validate emerging-corridor candidates against subsequent night-time-light observations.
  \item Incorporate road-network density and travel-time accessibility as additional corridor-structure dimensions.
  \item Extend corridor-object analysis to named transport axes (e.g., Lagos--Ibadan, Aba--Onitsha, Kano--Kaduna--Zaria).
\end{itemize}

\section{Conclusion}
\label{sec:conclusion}

This paper has demonstrated a reproducible, proxy-based framework for screening economic-corridor candidates from multi-source earth observation and commercial geospatial data at cell-level resolution, applied across Nigeria's kilometre-scale analytical frame (762,751 cells) using the 749,191-cell national-grid subset for national, state, and zone summaries.

The main findings are:

\begin{enumerate}[leftmargin=2em]
  \item \textbf{11,243 active corridor cells} (1.50\% of the national-grid subset) pass the dual proxy gate.
  \item \textbf{598 corridor objects} are constructed by clustering adjacent active cells, spanning 10,682 cells and approximately 36.3 million people, classified into five substantive typologies plus a Minor Cluster residual: Mature Urban Cores (5), Growth Corridors (133), Commercial Axes (113), Emerging Corridors (217), and Constrained Corridors (119).
  \item \textbf{Metro corridor density} is highest in Benin City (58.1\%), followed by Port Harcourt (45.9\%), Kaduna (39.2\%), Ibadan (37.1\%), and Lagos (36.0\%); Lagos has the most corridor cells in absolute terms (1,193) but ranks fifth in density over its larger configured footprint.
  \item A \textbf{south--north gradient} in corridor density persists across geopolitical zones, from the South East (6.19\%) to the North East (0.36\%).
  \item \textbf{Facility point-presence context} from 46,146 GRID3 health facilities and 204,513 schools accompanies the corridor objects; these are point counts, not measures of service access, capacity, or functionality.
\end{enumerate}

The corridor score ($0.30 \times \text{builtup} + 0.25 \times \text{nl\_norm} + 0.20 \times \text{UII} + 0.15 \times \text{bldg\_norm} + 0.10 \times \text{smb\_norm}$) provides a transparent, reproducible, and extensible screening framework for corridor detection in data-scarce environments. It identifies candidate corridor cells and objects, reveals metro, state, and zone patterns, and supplies screening intelligence for spatial planning, infrastructure prioritisation, and further validation. It does not measure actual GDP, firm revenue, transaction volume, freight movement, infrastructure capacity, service adequacy, or official corridor designation. Future work should validate corridor candidates against external economic records and multi-year dynamics. Public-safe tables, methodology notes, and code/index definitions are released through the project GitHub repository, with third-party source datasets governed by their respective licences (Section~\ref{sec:availability}).

\bigskip
\noindent\textbf{Correspondence:} \texttt{research@hamoye.org}

\section{Data and Code Availability}
\label{sec:availability}

Public-safe derived tables, methodology notes, and index definitions are made available through the project GitHub repository: \url{https://github.com/RichardRe2/Hamoye/tree/main/research}. The repository contains derived outputs and reproducibility materials sufficient to inspect the corridor-score construction, the dual proxy gate, the typology definitions, and the public tables reproduced in this paper (corridor objects; metro, state, and zone corridor density; constrained and emerging candidate cells; and index-component definitions).

Third-party source datasets --- including ESA WorldCover, NASA VIIRS Black Marble, Google Open Buildings, WorldPop, OpenStreetMap and Overture Maps, GRID3 facilities, and geoBoundaries --- remain governed by their original licences and are not redistributed unless permitted; the repository releases only derived, public-safe outputs. Questions regarding data access, reproducibility, or future programmatic access should be directed to \texttt{research@hamoye.org}.

\medskip
\noindent The corridor index has not been validated against firm-level revenue, tax records, freight flows, or official business-register data. All corridor outputs are proxy-based screening signals and should be interpreted accordingly.

\bibliography{references}

\end{document}